\documentstyle[11pt,psfig,newpasp,twoside]{article}
\index{supernova remnants}
\index{pulsar wind nebulae}
\index{X-rays}
\index{Chandra X-ray Observatory}
\index{jets and outflows}
\index{pulsars}
\index{PSR~B1509--58}
\index{SNR~G320.4--1.2}
\index{MSH~15--5{\em 2}}
\index{RCW~89}

\def\cxo{{\em Chandra}}
\def\psr{B1509--58}
\def\snr{G320.4--1.2}
\def\HI{H\,{\sc i}}
\def\etal{{et~al.}}

\def\edcomment#1{\iffalse\marginpar{\raggedright\sl#1\/}\else\relax\fi}
\reversemarginpar

\begin{document}
\title{Chandra Observations of Pulsar \psr\ and Supernova Remnant \snr}
\author{Bryan M. Gaensler\altaffilmark{1}}
\affil{Center for Space Research, Massachusetts Institute of Technology,
Cambridge MA, USA}
\author{Jonathan Arons, Michael J. Pivovaroff}
\affil{University of California, Berkeley CA, USA}
\author{Victoria M. Kaspi}
\affil{McGill University, Montreal QC, Canada}

\altaffiltext{1}{Current address: 
Harvard-Smithsonian Center for Astrophysics, Cambridge MA, USA}

\begin{abstract}

The young and energetic pulsar \psr\ powers a bright X-ray synchrotron
nebula, embedded in the unusual supernova remnant \snr.  We present
observations of this system with the {\em Chandra X-ray Observatory},
which show a spectacularly complicated source. The nebula is dominated
by a bright collimated feature which we interpret as a relativistic jet
directed along the pulsar spin axis. Several compact knots can be seen
in the immediate vicinity of the pulsar. While many of these features
are similar to those seen around the Crab pulsar, the nebula
surrounding PSR~\psr\ shows important differences which are possibly a
result of the the latter's low nebular magnetic field and low density
environment.

\end{abstract}

\section{Introduction}

\begin{figure}
\centerline{\psfig{file=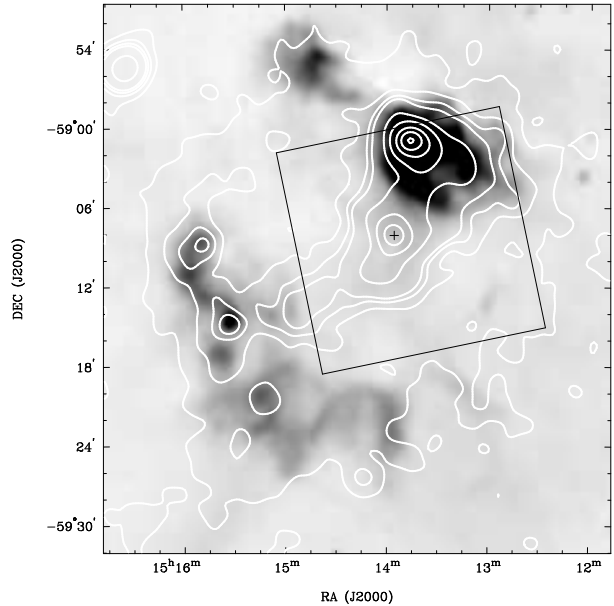,height=8cm,angle=270}}
\caption{A radio/X-ray comparison of \snr. 
The greyscale corresponds
to 843-MHz MOST observations, while
the white contours represent smoothed {\em ROSAT}\ PSPC data.
The position of PSR~\psr\ is marked
with a ``+'', while the black box delineates the ACIS-I field-of-view.}
\label{fig_1}
\end{figure}

The supernova remnant (SNR) \snr\ (MSH~15--5{\em 2}) has unusual
radio and X-ray properties. At radio wavelengths (Figure~\ref{fig_1}),
its morphology is dominated by two distinct limbs of radio emission.
Superimposed on the northern limb is a bright core of emission,
coincident with the H$\alpha$ nebula RCW~89. The distinctly
non-circular appearance of this source has been attributed
to expansion into an elongated cavity, a claim supported
by recent \HI\ observations of this region (Dubner \etal\ 2002). 
\HI\ absorption
towards the SNR demonstrates it to be at a distance of 5~kpc (Gaensler
\etal\ 1999).

At X-ray energies (Figure~\ref{fig_1}), the system is dominated by a bright
central point source. This corresponds to the pulsar \psr,
which has also been detected at radio wavelengths and in $\gamma$-rays.
PSR~\psr\ is one of the youngest and most energetic pulsars known:
it has a spin-period $P=151$~ms, a magnetic field
$B = 1.5\times10^{13}$~G, a spin-down luminosity $\dot{E} =
1.8\times10^{37}$~erg~s$^{-1}$ and a characteristic age $\tau = 1700$~yr.
Surrounding the pulsar is an elongated non-thermal nebula, presumed
to be the pulsar wind nebula (PWN) powered by the pulsar's spin-down;
no radio counterpart to this PWN has been identified. To the north
of PSR~\psr\ is a source of thermal X-rays, coincident with the
bright radio and optical emission from RCW~89
(Trussoni \etal\ 1996; Tamura \etal\ 1996).

Existing observations have raised a number of issues regarding
this pulsar and its interaction with its environment. Firstly,
the elongated morphology of the PWN suggests that its morphology
is dominated by a collimated outflow directed along the pulsar
spin-axis, which possibly collides with and is interacting with the RCW~89
region (Manchester \& Durdin 1983; Tamura \etal\ 1996; Brazier
\& Becker 1997; Gaensler \etal\ 1999).
Greiveldinger \etal\ (1995) have claimed that there is a compact
disc of nebular emission immediately surrounding the pulsar,
while Brazier \& Becker (1997) rather propose a ``cross''-shaped
morphology in this region, which they interpret as an equatorial torus
and polar jets, seen edge-on.

Clearly our understanding of this complicated source
can benefit from observations at higher angular resolution.
We have consequently carried out observations of PSR~\psr\
and its surroundings with the {\em Chandra X-ray Observatory}.
We summarize these results below; these data are discussed in
more detail by Gaensler \etal\ (2002).

\section{X-ray Observations}

\subsection{Imaging}

\begin{figure}
\centerline{\psfig{file=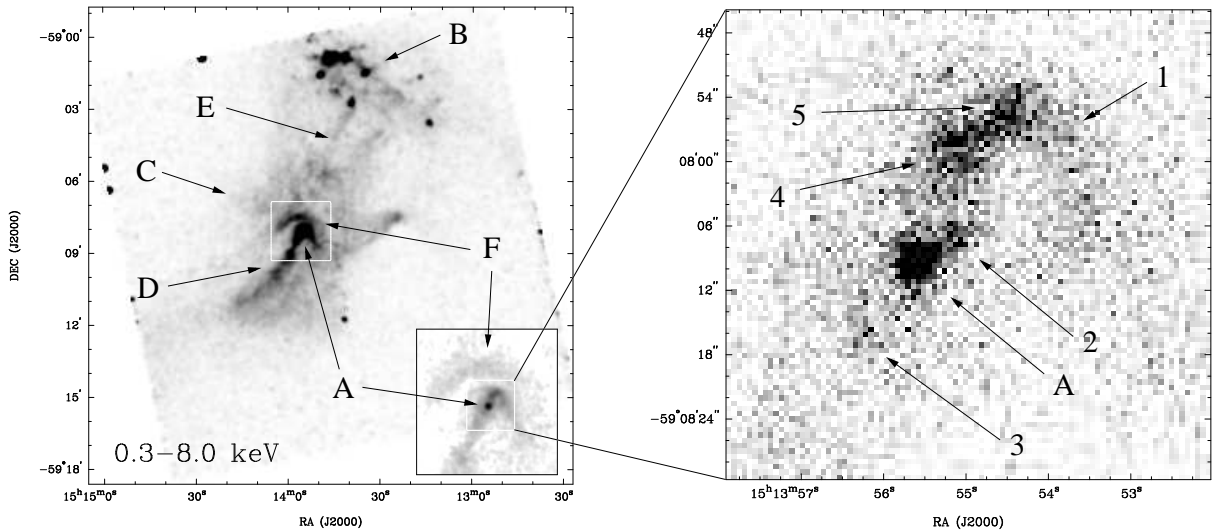,height=7cm}}
\caption{\cxo\ images of PSR~\psr / SNR~\snr\ in the energy range
0.3--8.0~keV. The left-hand panel shows the entire field, 
exposure-corrected and convolved with a $10''$ gaussian. The white
box in the main image shows the region covered by the inset at lower right.
The right-hand panel shows the region immediately surrounding the
pulsar. Specific features discussed in the text are indicated.}
\label{fig_2}
\end{figure}

PSR~\psr\ was observed with the ACIS-I detector on \cxo\ on 
2000~Aug~14, with an effective exposure time of 19~ks. The resulting
image is shown in the left panel of
Figure~\ref{fig_2}. A number of features can be seen
in this image: the pulsar itself (marked as A),
a ring of X-ray clumps coincident with RCW~89 (B),
a diffuse elongated PWN surrounding the pulsar (C),
a ``jet'' feature lying along the PWN's main axis (D),
a possible counterpart to the ``jet'' seen as
a region of {\em reduced}\ emission (E), and
an arc just to the north of the pulsar which is bisected by the main nebular
axis (F).

The right panel 
of Figure~\ref{fig_2} shows the region immediately surrounding the pulsar
at the full resolution of the data. This Figure shows an inner
arc sitting inside feature F (marked as 1), and
several knots close to the pulsar (2, 3, 4, 5).

\subsection{Spectroscopy}

We have extracted over 50\,000 counts from the diffuse
PWN (feature C in Figure~\ref{fig_2}); these data
show no spectral features. Fitting them
with a power law, we find an absorbing column
$N_H = (9.5\pm0.3)\times10^{21}$~cm$^{-2}$ and a photon
index $\Gamma = 2.05\pm0.04$, in good agreement with previous
results. The brighter features superimposed on this
diffuse emission all have distinctly harder spectra:
the ``jet'' (feature~D), outer arc (F) and inner
arc (1)
all have $\Gamma \approx 1.6-1.7$, while
the innermost knots (2--5) all have $\Gamma \approx 1.3-1.6$.

The spectrum of the RCW~89 region shows several clear
emission lines, confirming its interpretation as a thermal source
distinct from the other components of the system (e.g.\ Tamura \etal\ 1996).
Preliminary analysis of these data shows that the emission 
can be approximately fit by a non-equilibrium ionization model
with variable abundances.

\section{Discussion}

\subsection{Radio/X-ray Comparison}

\begin{figure}
\centerline{\psfig{file=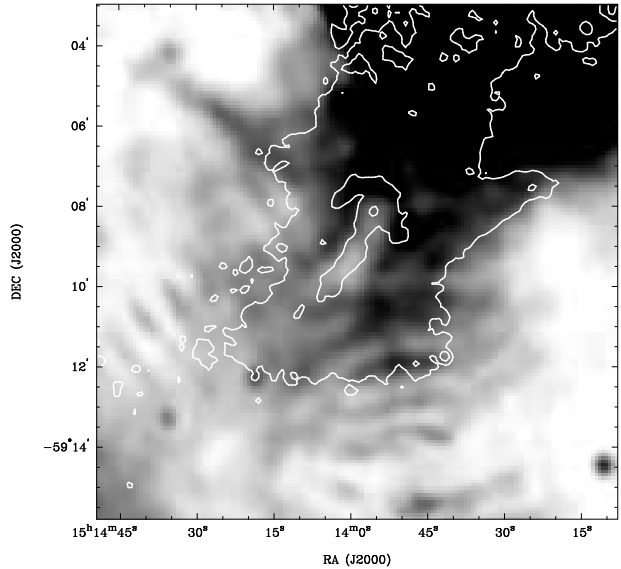,height=7.5cm,angle=270}}
\caption{Radio and X-ray emission in the region
surrounding PSR~psr. The greyscale 
shows radio data from the 1.4-GHz observations
of Gaensler \etal\ (1999), whil the contours show \cxo\ data as
in Figure~\ref{fig_2},
with contours at 0.5\%, 2\% and 30\% of the peak.}
\label{fig_3}
\end{figure}

As shown in Figure~\ref{fig_3}, diffuse radio emission seen
in the vicinity of the pulsar
closely matches the extent of the diffuse X-ray PWN
(feature~C in Figure~\ref{fig_2}). It thus seems clear that this radio
emission is the
long-sought radio PWN.  An elongated region of reduced radio emission
can be seen to the south of the pulsar,
corresponding closely to the morphology of the X-ray
``jet'' (feature~D).  This argues that this is a physically distinct
structure  within the nebula, and does not simply result from
variations in brightness.

\subsection{The Diffuse Nebula}

The X-ray emission surrounding \psr\ shows a clear
symmetry axis, oriented at a position angle
$150^\circ \pm 5^\circ$ (N through E), and
manifested on all spatial scales between $10''=0.2$~pc
and $10'=15$~pc. Such alignment can only be
enforced by the central pulsar. We think it 
likely that this axis represents the spin-axis of
the pulsar, as has also been argued for the Crab and Vela pulsars 
(Hester \etal\ 1995; Helfand, Gotthelf \& Halpern 2001).

A variety of different arguments imply that the magnetic
field in the PWN is approximately 8~$\mu$G (see Gaensler
\etal\ 2002 for details). This is
an order of magnitude weaker than in the Crab Nebula,
and implies a spectral break due to synchrotron cooling
just below the X-ray band. This low magnetic field is
most likely a result of the low density of the medium
into which the PWN has expanded (Bhattacharya 1990).
It is unlikely
the pulsar is much older than $\tau$ (e.g.\ Gvaramadze,
these proceedings), as this would require an even lower
nebular field strength, not consistent with other estimates.

\subsection{Outflow and Orientation}

Feature~D has a distinctly harder synchrotron spectrum than the surrounding
PWN. If the injected electron spectrum in the two regions
is the same and the spectral difference between them is due
to synchrotron cooling, it can be shown that the flow velocity in feature~D
must be $>0.2c$ (Gaensler \etal\ 2002). Thus this source
corresponds to a true jet, which we have calculated carries away
$>0.5\%$ of $\dot{E}$. This jet is directed along the pulsar
spin-axis, and appears to be a much larger and more spectacular
version of similar features seen for the Crab and Vela PWNe 
(Hester \etal\ 1995; Helfand \etal\ 2001).

While no direct counterpart to this outflow can be seen
to the north of the pulsar, the collimated nature of
feature~E suggests that we are seeing a cylindrical sheath
of emission around an unseen counter-jet. This lack of
direct emission can be simply explained by Doppler boosting,
provided the jet axis is inclined by $\zeta \la30^\circ$ to the
line-of-sight. This is contrary to the
edge-on morphology, $\zeta \sim 70^\circ$, argued
by Brazier \& Becker (1997) from lower-resolution data.

\subsection{Arcs and Inner Structure}

Feature~F is distinctly one-sided, and so can potentially be
interpreted as a bow-shock driven by the pulsar.
However, this feature is embedded within a much larger PWN,
in which the pulsar's velocity cannot be supersonic.

We thus prefer to interpret this arc as a toroidal
structure, lying in a plane perpendicular to the main
symmetry axis. In this case,
the projected morphology of the arc implies $\zeta < 30^\circ$,
consistent with the estimate made above from Doppler boosting
of the jet.

Considered together, the inner and outer arcs (features~1
and F respectively) resemble the inner ring and outer torus
seen in \cxo\ images of the Crab Nebula (Weisskopf \etal\ 2000;
Mori, these proceedings). There are
two characteristic time-scales associated with such
structures: the time-scale for radiative losses via
synchrotron emission, and the flow time from the pulsar.
In the torus of the Crab Nebula, these times
are comparable. However, for PSR~\psr\ the radiative time-scale
is $\sim30$ times longer than the flow time due to the lower
nebular magnetic field. Thus the arcs seen here must
be dynamical, not radiative, features in the PWN. We show elsewhere
that they can be interpreted as ``wisps'' as seen in the
Crab Nebula, resulting from ion compression in a particle-dominated flow
(Arons, these proceedings; Gaensler \etal\ 2002). In
this case, we might expect outward motion of these features
as is seen for the Crab. The expected proper motion would
be a few arcsec per year, easily detectable with \cxo.

An interpretation for the knots close to the pulsar (features 2--5)
is less clear. 
These knots may be emission from
the unshocked pulsar wind, analogous to ``knot~1''
and ``knot~2'' seen with {\em HST}\ close to the Crab pulsar (Hester
\etal\ 1995).

\section{Conclusion and Questions}

These \cxo\ data have provided a wealth of new information on
PSR~\psr\ and its interaction with its environment.  We have confirmed
the presence of a collimated flow directed along the pulsar spin-axis,
and have argued that the flow is relativistic and is inclined at
$<30^\circ$ to our line-of-sight.  We have interpreted two arcs of
emission seen close to the pulsar as dynamical features in an
equatorial flow, and have identified several knots at separations
$<0.5$~pc from the pulsar.

Many issues still need to be investigated. Do the arcs
and knots show motion and/or variability? Do any of
these features have counterparts at other wavelengths?
What is the nature of RCW~89 and its thermal clumps?
While these questions still remain, it is clear that
PSR~\psr\ provides a new opportunity to probe the detailed
structure of a pulsar wind.

\acknowledgments

This work was supported by NASA through SAO grant
GO0-1134X, Hubble Fellowship grant HST-HF-01107.01-A (B.M.G.), 
contracts NAS8-37716 and NAS8-38252 (M.J.P.), and
LTSA grant NAG5-8063 (V.M.K.).

\end{document}